\documentstyle{article}
\begin{document}

\title{{\bf Time evolution of the Partridge-Barton Model}} 
\author{Roberto N. Onody\thanks{Eletronic address:onody@ifsc.sc.usp.br} and
Nazareno G. F. de Medeiros \thanks{Eletronic address:ngfm@ifsc.sc.usp.br}\\
{\small {\em Departamento de F\'{\i}sica e Inform\'{a}tica } }\\
{\small {\em Instituto de F\'{\i}sica de S\~{a}o Carlos} }\\
{\small {\em Universidade de S\~{a}o Paulo - Caixa Postal 369} }\\ 
{\small {\em 13560-970 - S\~{a}o Carlos, S\~{a}o Paulo, Brasil.}}}
\date{} 
\maketitle 
\normalsize 
\baselineskip=16pt 
\hyphenation{pro-ba-bi-li-ties}

\begin{abstract}

The time evolution of the Partridge-Barton model in the presence
of the pleiotropic constraint and deleterious somatic mutations 
is exactly solved for arbitrary fecundity in the context of a 
matricial formalism. Analytical expressions for the time dependence
of the mean survival probabilities are derived. Using the
fact that the asymptotic behavior for large time $t$ is controlled
by the largest matrix eigenvalue, we obtain the steady state values
for the mean survival probabilities and the Malthusian growth exponent.
The mean age of the population exhibits a $t^{-1}$ power law decayment.
Some Monte Carlo simulations were also performed and they corroborated 
our theoretical results.

\vspace{3cm}


\end{abstract}

\newpage

\section{\bf Introduction}

\indent

Early in life we perceive that everything around us, inanimate objects,
animals and human beings undergo a variety of changes that accompany the
passage of time. Everything suffers a progressive deterioration
with time. This phenomenon is called ageing or senescence and it is
characterized by a decline in the physical capabilities of the
individuals. Several theories (see \cite{bern} and references therein)
have been suggested to explain why
there is senescence, when it occurs and what are the biological processes 
responsible for it. Usually, these theories are divided into three classes:
biochemical, evolutionary and telomeric. The first invokes damages
on DNA, cells, tissues and organs and connect senescence with imperfections of
the biochemical processes. One kind of this biochemical imperfection is the 
presence of free radicals which can cause death of the cells or may even 
lead to cancer \cite{kirk}. The evolutionary theory \cite{rose,worth}, 
on the other hand,  
explains the senescence as a competitive result of the reproductive rate,
mutation, heredity and natural selection. In the telomere hypothesis \cite{red}, 
senescence depends on the cumulative number of cell divisions. The 
replication of a normal cell is followed by a telomeric shortening. This
acts as a counting mechanism which controls the number of divisions.  

Evolutionary theories of ageing are hypothetico-deductive in character,
not inductive. They do not contain any specific genetic parameter, but only
physiological factors and constraints imposed by the environment. There are 
two kinds:
the optimality theory and the mutational theory. In the optimality theory
\cite{topt}, senescence is a result of searching an optimal life history where
survival late in life is sacrificed for the sake of early reproduction.
For the mutational theory \cite{worth,tmut}, on the other hand, ageing is a
process which comes from a balance between Darwinian selection and
accumulation of mutations. The natural selection efficiency to remove harmful
alleles in a population depends on when in the lifespan they come to express. 
Alleles responsible for lethal diseases that express late in life, escape
from the natural selection and accumulate in the population, provoking  
senescence. Nevertheless, if the natural selection is too strong then
deleterious mutations might not accumulate in the population and the eternal
youth could be reached. An evolutionary model with such characteristics
was recently studied and solved by Onody and de Medeiros \cite{onody}.
 
A simple evolutionary model of ageing is the Partridge-Barton model\cite{pb}.
It was introduced to illustrate the optimality theories of ageing. Its 
principal feature is the inclusion of the antagonistic balance mechanism 
\cite{antg}. This mechanism arises out from processes which
enhance the lifespan early in life, but have deleterious effects latter. 

In this work, we find an exact solution for the whole dynamics of the
Partridge-Barton model. When only deleterious somatic mutations and pleiotropy
are present the time evolution of the model can be formulated in a matricial 
form. Explicit analytic expressions can be written for the mean survival
probabilities and the growth rate. For large time $t$, the system behavior
is dominated by the matrix largest eigenvalue. The existent integrals
can be solved by the saddle point approximation, allowing us to determine
precisely the steady state values of the survival probabilities. A time
expansion for the population's mean age shows that it converges to a constant
value according to a $t^{-1}$ power law, a result which was first obtained 
by Ray \cite{ray}. All the results were confirmed by some Monte Carlo 
simulations that we performed.

\section{\bf The Partridge-Barton Model}

\indent

In the Partridge-Barton model there are only three ages. The population consists
of babies $(age=0)$, juveniles $(age=1)$ and adults $(age=2)$. The survival 
probabilities from infancy to juvenile is $J_{1}$ and from juvenile to adulthood 
is $J_{2}$. Reproduction is permitted only to juveniles and adults, with rates 
$m_{1}$ and $m_{2}$, respectively. Babies don't have
offsprings and adults are eliminated from the population after reproduction. 

The population  grows at a steady rate
$r$. The Malthusian growth exponent $r$ is related to the other parameters 
of the model through a discrete version of the
Euler-Lotka equation \cite{mur}
\begin{equation}
m_{1}J_{1}e^{-r}+m_{2}J_{1}J_{2}e^{-2r}=1.
\end{equation}

The antagonistic pleiotropy \cite{antg} arises when the same gene is responsible
for multiple effects. For example, genes enhancing early survival by promotion
of bone hardening might reduce later survival by promoting arterial hardening.
Partridge and Barton implemented the basic ideas of the antagonistic 
pleiotropy by adopting the 
constraint, $J_{1}+J_{2}^{x}=1$, between the survival probabilities $J_{1}$ and 
$J_{2}$. The
parameter $x$ is a real positive number whose value depends on the kind of 
population we are dealing with. 
The pleiotropic condition states that it is impossible to sustain simultaneously
both high juvenile and adult survivals. 
For the particular case in which $m_{1}=m_{2}=1$ and $x=4$, 
Partridge and Barton found $J_{1}=0.935$ and $J_{2}=0.505$ as the values which 
maximize the growth rate $r$.

Also the action of deleterious or helpful mutations can be added to the model.
Using Monte Carlo simulations, Stauffer \cite{stf} studied the case in which
the pleitropic constraint $J_{1}+J_{2}^{4}=1$ is accompanied by random somatic 
mutations. His results clearly show that the survival 
probabilities $J_{1}$ and $J_{2}$ move rapidly to stationary values with 
$J_{1}>J_{2}$. This 
fact means that the model exhibits senescence, in the sense that the adult 
survival is lower than the juvenile. In the absence of 
mutations, $J_{1}$ and $J_{2}$ tend towards $0.935$ and $0.505$ in accord with 
the Partridge-Barton
conclusions. However, it is not clear how the system drives itself towards 
these optimal values.

\section{\bf Analytical Solution}

\indent
  
In this section we obtain the exact time solution of the Partridge-Barton model
in the presence of pleiotropy and somatic mutations. 

Let $N_i(J_i,t)$ be 
the number of individuals at age $i$ ($i=0,1,2$)
with survival probability between $J_i$ and $J_i+dJ_i$ at time $t$. We choose, 
as initial condition, a population with the profile

\begin{equation}
N_i(J_i,0) = N_0\delta_{i,0},
\end{equation}
that is, in $t=0$ there are only $N_0$ babies with the survival 
probabilities $J_0$ uniformly distributed in the interval $[0,1]$.

At time $t$, all babies are equally submitted to somatic and deleterious 
mutations with strength $\alpha$ ($\alpha < 1$). Their survival probabilities 
$J_{0}$ are changed to $J_1=\alpha J_{0}$. 
Subsequently, all these babies pass through natural selection in a such way
that, on average, the number of juveniles with survival probability $J_{1}$ 
at the instant $t+1$ is given by
\begin{equation}
N_{1}(J_{1},t+1)= J_{1}N_{0}(J_{0},t).
\end{equation}
Since the mutation is restricted to be {\em somatic}, 
each one of the $N_{1}(J_{1},t+1)$
juveniles will give birth to exactly $m_{1}$ offspring with survival 
probability $J_{0}$.

Now, the probability with which a juvenile will reach adulthood must
take into account the antagonistic pleiotropy and the somatic
deleterious mutations. As pleiotropy is not affected by the somatic mutations,
a juvenile with survival probability $J_{1}$ (formerly, a baby with survival 
probability $J_{0}$) will change its survival probability to $(1-J_{0})^{1/x}$,
where $x$ is a real positive number and a measurement of the pleiotropic 
constraint. Under the action of a deleterious somatic 
mutation, described by a parameter $\beta$ ($\beta < 1$, fixed), the new 
survival probability can be written as $J_{2}=\beta (1-J_{0})^{1/x}$.
Submitting all juveniles to natural selection we get, on average, the number
of adults with survival probability $J_{2}$ which is given by
\begin{equation}
N_{2}(J_{2},t+1) = J_{2}N_{1}(J_{1},t). 
\end{equation}
Each one of these adults will generate $m_{2}$ descendants with survival
probability $J_{0}$ since the mutations are {\em not inherited}.

In general, the
number of babies with survival probability $J_0$ is given by 
\begin{equation}
N_0(J_0,t)=m_1N_1(J_1,t)+m_2N_2(J_2,t), \;\;\; for \; t \geq 1 
\end{equation}
where $J_1=\alpha J_{0}$ and $J_{2}=\beta (1-J_{0})^{1/x}$. If we substitute 
equation (5) into (3) we can write the following recursive matricial equation
\begin{equation}
\left( 
\begin{array}{c}
N_{1}(J_{1},t+1) \\ 
N_{2}(J_{2},t+1)
\end{array}
\right) = A \left(
\begin{array}{c}
N_{1}(J_{1},t) \\ 
N_{2}(J_{2},t)
\end{array}
\right), 
\end{equation}
where $A$ is the matrix
\begin{eqnarray*}
A=\left( 
\begin{array}{cc}
m_{1}J_{1} & m_{2}J_{1} \\ 
J_{2} & 0
\end{array}
\right).
\end{eqnarray*}
Iterating the equation above and using the initial condition, we get for
$t \geq 2$
\begin{equation}
\left( 
\begin{array}{c}
N_{1}(J_{1},t) \\ 
N_{2}(J_{2},t)
\end{array}
\right) =J_{1}N_{0}(J_{0},0)A^{t-2}\left( 
\begin{array}{c}
m_{1}J_{1} \\ 
J_{2}
\end{array}
\right), 
\end{equation}
with $A^{0}$ meaning the identity matrix.

The complete dynamics of the Partridge-Barton model can be obtained by
diagonalizing the matrix $A$. We have, explicitly (for $t \geq 2$)
\begin{eqnarray}
N_{1}(J_{1},t) &=&\frac{J_{1}N_{0}(J_{0},0)}{\sqrt{m_{1}^{2}J^{2}_{1}+
4m_{2}J_{1}J_{2}}}\left[m_{1}J_{1}\left(\lambda _{+}^{t-1}-\lambda _{-}^
{t-1}\right) \right. \nonumber \\
&&\left. +m_{2}J_{1}J_{2}\left(\lambda_{+}^{t-2}-\lambda _{-}^{t-2}\right) 
\right],
\end{eqnarray}
\begin{eqnarray}
N_{2}(J_{2},t) &=&\frac{J_{1}N_{0}(J_{0},0)}{\sqrt{m_{1}^{2}J^{2}_{1}+
4m_{2}J_{1}J_{2}}}\left[m_{1}J_{1}J_{2}\left( \lambda _{+}^{t-2}-\lambda_{-}^
{t-2}\right) \right.\nonumber  \\
&&\left. +m_{2}J_{1}J^{2}_{2}\left( \lambda_{+}^{t-3}-\lambda _{-}^{t-3}\right)
\right],
\end{eqnarray}
where 
\begin{equation}
\lambda _{\pm}=\frac{m_{1}J_{1}\pm \sqrt{m_{1}^{2}J^{2}_{1}+
4m_{2}J_{1}J_{2}}}{2}
\end{equation}
are the eigenvalues of the matrix $A$, $J_1=\alpha J_{0}$ and $J_{2}=\beta 
(1-J_{0})^{1/x}$. Let us point out that the time evolution 
of the babies distribution $N_0 (J_0,t)$, can be calculated using equations (5),
(8) and (9). Having the expressions above, we can determine the evolution of 
many other quantities like the total number of inviduals at age $i$ 
$N_{i}(t)=\int_{0}^{1} N_{i}(J_{i},t)dJ_{i}$ or their {\em mean survival 
probabilities}
$\langle J_{i}\rangle(t)=\frac{\int_{0}^{1}J_{i}N_{i}(J_{i},t)dJ_{i}}
{\int_{0}^{1}N_{i}(J_{i},t)dJ_{i}}$. The given {\em input parameters } are
the initial population $N_0$, the birth rates ($m_1$ and 
$m_2$), the mutation strengths ($\alpha$ and $\beta$) and the pleiotropic 
constraint ($x$).

\section{\bf Asymptotic Limit}

\indent

Before taking the asymptotic limit, we observe that $\lambda_+$
is the largest eigenvalue for {\em all} possible values of the input parameters.
Once these parameters are fixed and 
$J_{2}=\beta (1-J_{1}/ \alpha)^{1/x}$, $\lambda_+$ is in the last instance a 
function of $J_1$. From the equation ($8$) we have, asymptotically

\begin{equation}
N_{1}(J_{1},t)  \approx e^{t \; ln[\lambda_{+}(J_{1})]}.
\end{equation}

By integrating in $J_1$ the expression above, we can get the total number of 
juveniles $N_1(t)$. 
It is convenient to change the integration variable $J_1$ for a new variable $y$
(a monotonically increasing function of $J_1$), $y=-cot(\pi J_{1})$, such that

\begin{equation}
N_{1}(t) \approx \int_{-\infty}^{\infty}\frac{e^{t \; ln[\lambda_{+}(y)]}}
{\pi(1+y^2)}dy.
\end{equation}

For large time $t$, this integral can be evaluated by the saddle point 
approximation. We thus obtain

\begin{equation}
N_{1}(t) = A(\tilde{y}) \frac{e^{t \; ln[\lambda_{+}(y)]}}
{\sqrt{t}},
\end{equation}
where $\tilde{y}$ is the value which maximize the eigenvalue $\lambda_{+}$ and
$A(\tilde{y}) = \sqrt{\frac{\pi}{\frac{-1}{2 \lambda_+} \frac{
d^2 \lambda_+}{ dy^2}|_{y=
\tilde{y}}}}$. 

In the original paper of Partridge and Barton,
the optimization process was achieved by a direct (and not well explained)
maximization of the growth rate. Here, in our formalism, it is a simple and a
natural consequence of taking the asymptotic time limit in the exact evolving 
equations. Further, the growth rate or the Malthusian exponent is simply given 
by $ln[\lambda_{+}(\tilde{y})]$.

To have deepest insight in the dynamics, let us determine the probability
density $P_1(J_1,t)$ of finding a juvenile at time $t$ with survival 
probability between $J_1$ and $J_1 + dJ_1$. It is given by

\begin{equation}
P_{1}(J_1,t) = \frac{N_1(J_1,t)}{\int_0^1 N_1(J_1,t) dJ_1}=
\frac{N_1(J_1,t)}{N_1(t)} \approx \sqrt{t} \; e^{t \; ln[\frac{\lambda_+(J_1)}
{\lambda_+(\tilde{J_1})}]}
\end{equation}
where we have used equations ($11$) and ($13$) and 
$\tilde{y}=-cot(\pi \tilde{J_{1}})$. Clearly, at the asymptotic limit, the 
distribution probability $P_{1}(J_1,t \rightarrow \infty)$ approaches the 
Dirac delta function
$\delta (J_1- \tilde{J_1})$ and the mean survival probability at age $1$,
is simply given by
$ \langle J_1 \rangle = \tilde{J_1}$. Similar results can be obtained for the
ages $0$ and $2$. Another interesting quantity which can be calculated is
the population mean age $ \langle A \rangle (t)$ defined as
$ \langle A \rangle (t) = \frac{\sum_{i=0}^{2} i \; N_i(t)}
{\sum_{i=0}^{2} N_i(t)}$. It is straightforward to show that

\begin{equation}
\langle A \rangle (t) = \frac{\gamma + 2}{\gamma (1 + m_1) + (1+m_2)} +
\left\{ \frac{2 \gamma (1+m_1) + \gamma (1+m_2)}{2[\gamma (1 + m_1) + 
(1+m_2)]^2} \right\} t^{-1} + O(t^{-2})
\end{equation}
where $\gamma = \frac{\lambda_+ (\tilde{J_1})}{\tilde{J_2}}$ with 
$\tilde{J_{2}}=\beta (1-\tilde{J_{1}}/ \alpha)^{1/x}$. So we rederive, in a
quite simple way, the power law decayment first found by Ray \cite{ray}. 

\section{\bf Discussion}

\indent

We solved exactly in this paper the Partridge-Barton model under 
the action of arbitrary pleiotropic constraints and deleterious
somatic mutations. Through a matricial formalism we were able to 
predict the complete time evolution of the population. We derived
analytic expressions for the time dependence of the mean survival 
probabilities and the Malthusian exponent. Since for large time $t$
the system behavior is controlled by the largest eigenvalue, it was
possible to obtain the steady state values of the survival probabilities 
and to demonstrate, in a simple way, that the population mean age has
a power law $t^{-1}$ decayment to its final constant value.

For comparison with our analytical results, we also performed some 
Monte Carlo simulations. In these simulations, the natural selection is
implemented by discarding any individual with survival probability 
smaller than a random number (generated from a uniform distribution).
The deleterious somatic mutations and the antagonic pleiotropy can be easily
incorporated into the computer program. More difficult is to avoid an 
explosion of the computer's memory due to the unlimited growth of the 
population. To take this problem into account, we resort to the Verhulst 
factor \cite{mur} which is commonly used in such circumstances. 

In Figure 1 we put together the analytical solution and the Monte Carlo 
result. The exact solution was plotted by inserting equations ($8$, $9$ and 
$10$) into the expressions for the mean survival probabilities
$\langle J_{i}\rangle(t)$ and by integrating them using the software Maple 
\cite{maple}. We conclude that the Monte Carlo simulations confirm very well the 
theoretical results. 

Finally, let us to point out that, unfortunately, the technique 
developed here cannot be applied to the case in which mutations are hereditary.
The main reason for this come from the fact that the equation ($5$) is no 
longer valid.

\section {\bf Acknowledgements}

We acknowledge CNPq (Conselho Nacional de Desenvolvimento Cient\'{\i}fico e 
Tecnol\'ogico) for the financial support.

\newpage

\begin{center} 
FIGURE CAPTION 
\end{center}

\vspace{2.0cm}

Figure 1. The continuous lines correspond to the analytical solutions and the
square symbols to the Monte Carlo simulations. We used $\alpha = 0.82$,
$ \beta = 0.67 $ , $x =4$, $m_1 = m_2 =1$ and $N_0 = 4000 $. The steady state
values are $\tilde{J_1} = 0.77$ and $\tilde{J_2} = 0.33$. There is senescence,
i. e., $J_2 \; < \; J_1$.

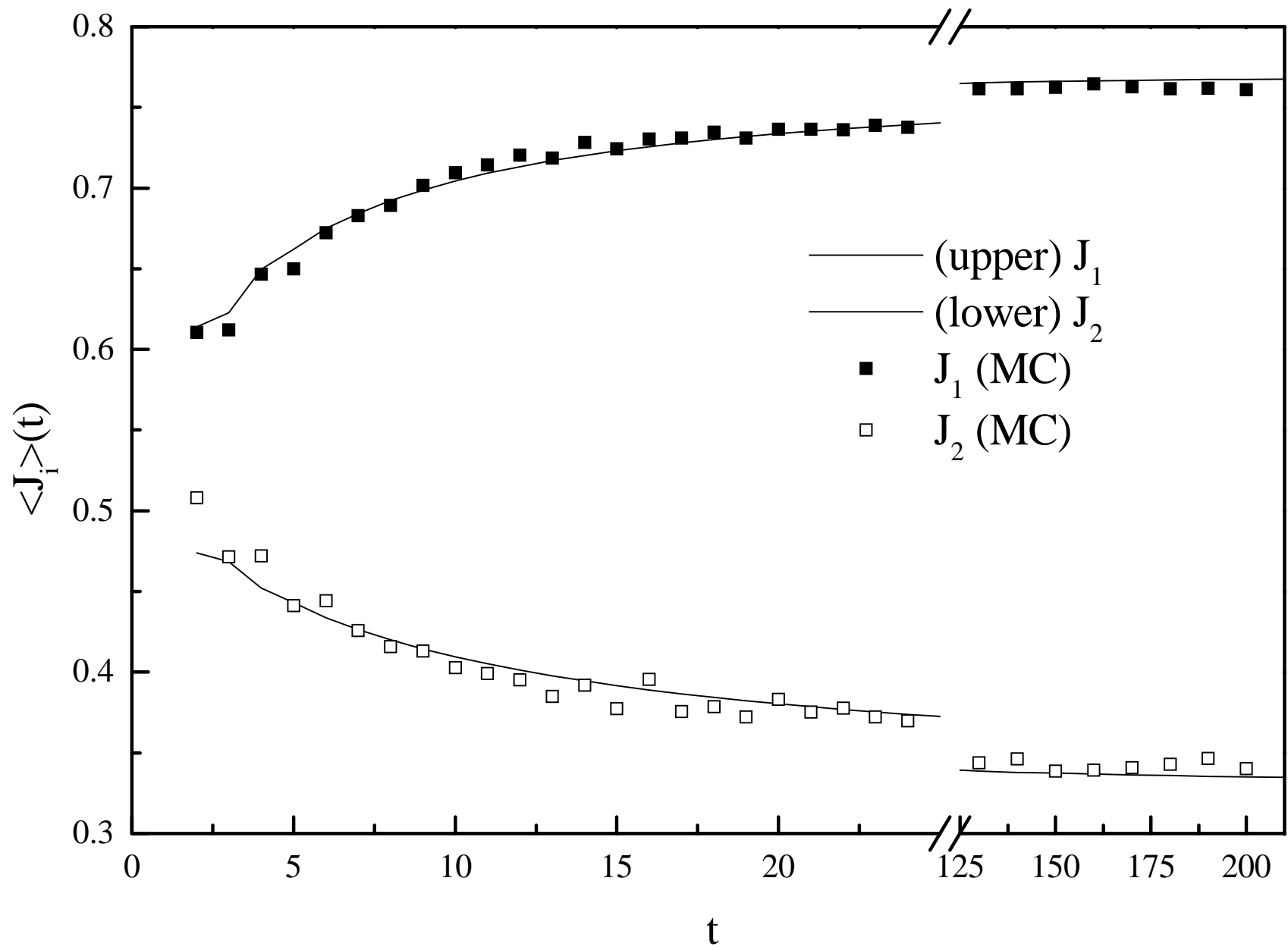
\end{document}